\documentclass[aps,prl,twocolumn,showpacs,groupedaddress,floatfix]{revtex4}

\usepackage{graphicx}
\usepackage{dcolumn}
\usepackage{epsfig}
\usepackage{amssymb}
\usepackage{amsmath}
\usepackage{rotating}
\usepackage{color}
\newcommand{\gammap}{\dot{\gamma}}
\newcommand{\kb}{k_{\scriptscriptstyle B}}
\newcommand{\chiA}{\chi_{\scriptscriptstyle I}}
\newcommand{\chiB}{\chi_{\scriptscriptstyle II}}
\newcommand{\AsA}{A_{\scriptscriptstyle I}}
\newcommand{\AsB}{A_{\scriptscriptstyle II}}
\newcommand{\taua}{\tau_{\scriptscriptstyle 1}}
\newcommand{\taub}{\tau_{\scriptscriptstyle 2}}

\begin{document}

\title{Observation of Droplet Size Oscillations in a Two-Phase Fluid under Shear Flow}

\author{Laurent Courbin}
\affiliation{Centre de Physique Optique Mol\'eculaire Hertzienne
UMR 5798, 351 Cours de la Lib\'eration, 33400 Talence, FRANCE}
\author{Jean-Baptiste Salmon}
\affiliation{Centre de Recherche Paul Pascal, Avenue Schweitzer, 33600 Pessac, FRANCE}
\author{Pascal Panizza}
\email{ppanizza@cribx1.u-bordeaux.fr} \affiliation{Centre de
Physique Optique Mol\'eculaire Hertzienne UMR 5798, 351 Cours de
la Lib\'eration, 33400 Talence, FRANCE}

\date{\today}
\begin{abstract}
Experimental observations of \textit{droplet size sustained oscillations} are reported in a two-phase flow between a
lamellar  and a sponge  phase. Under shear flow, this system presents two different steady states made of monodisperse
 multilamellar droplets, separated by a shear-thinning transition. At low and high shear rates, the droplet size
results from a balance between surface tension and viscous stress whereas for intermediate shear rates, it becomes
a \textit{periodic function of time}. A possible mechanism for such kind of oscillations is discussed.
\end{abstract}
 \maketitle

Homogenizing immiscible fluids by the use of shear flow is an
everyday experience  and a fundamental step in the processing of
soft materials \cite{Larson:99}. By fragmenting large domains, the
flow opposes the thermodynamics instability driving phase
separation and leads to the formation of \textit{non-equilibrium
steady states} in which the coarsening is stopped \cite{Onuki:97}.
Since the pioneering work of Taylor \cite{Taylor:34} on isolated
Newtonian emulsions, the study of the rupture and deformation of
isolated droplets under shear flow has drawn much attention. For
Newtonian fluids, the dispersed phase forms generally somewhat
deformed spherical droplets of the break-up size: $R \approx
\chi/\eta\gammap$, where $\chi$ is the surface tension between
both phases, $\eta$ the shear viscosity and $\gammap$ the shear
rate \cite{Taylor:34}. However, when the shear response of the
fluids exhibits complexity as in semi-dilute entangled polymer
solutions \cite{Hashimoto:95}, near critical mixtures
\cite{Min:89}, lamellar-sponge phase-separated mixtures
\cite{Cristobal:00}, or emulsions \cite{Mason:96}, other  steady
state structures such as cylindrical domains (known as the string
phase), two dimensional domains (ribbons phase) or colloidal
crystals made of droplets are observed. Predicting the steady
state domain morphology which results from the interplay between
the flow, the surface tension, the volume fraction of the two
phases and their viscosities is a real challenge in
non-equilibrium physics. Moreover, for a wide class of complex
fluids, several different structures can be obtained under shear
flow; these steady states are thus separated by out-of-equilibrium
transitions. Depending on the nature of the rheological transition
(shear-thinning or shear-thickening) and on the imposed dynamic
variable (stress or shear rate), structural bistability and/or
coexistence between structures characterized by shear-banding, can
be seen \cite{Cristobal:00,Bonn:98,Olmsted:97}. Because energy is
constantly brought into the system, these transitions are also
expected to lead to richer behavior such as bifurcations to
oscillatory states or even to chaos
\cite{Wunenburger:01,Grosso:01,Bandyopadhyay:00,Head:01,Picard:02}.

The purpose of this letter is to show for the first time on
two-phase flow that the droplet size can be a \textit{periodic
function of time}. Those oscillations are {\it asymptotic} ({\it
i.e.} they do not correspond to a transient regime) and display a
huge period of time, typically a few thousands of seconds.

We study a pseudo binary mixture made of sodium bis(2-ethylhexyl)
sulfo-succinate (AOT) and brine (water and sodium chloride)
\cite{Gosh:87}. At $T=25^\circ$C, for low salinities ($S \leq
14$~g.l$^{-1}$), flat bilayers of AOT, stack upon each other and a
lamellar phase ($L_\alpha$) is observed. At high salinities ($S
\geq 20$~g.l$^{-1}$), the bilayers interconnect randomly and
result in a Newtonian bicontinuous phase referred to in Literature
as sponge phase (or $L_{\scriptscriptstyle 3}$ phase)
\cite{Porte:92}. For intermediate salinities, coexistence between
$L_\alpha$ and $L_{\scriptscriptstyle 3}$ phases is found. We
prepare solutions in the two-phase
($L_\alpha/L_{\scriptscriptstyle 3}$) region with 20\%~wt. AOT
(from Fluka) and 80\%~wt. brine ($S=17$~g.l$^{-1}$) and let them
rest for a few weeks to reach equilibrium. Because of a density
mismatch between the two phases, an interface appears and the
volume fraction of $L_\alpha$ phase (65\% at $T=25^\circ$C) can be
measured directly from the observation of the position of this
interface.

\textit{Experiments}: As previously described \cite{Courbin:02},
Small Angle static Light Scattering experiments (SALS) are
performed at imposed shear rate with a home made transparent
Couette cell, with gap $e=1$~mm and inner radius $R_i=16$~mm. In
short, a circularly polarized laser beam (He-Ne,
$\lambda=632.8$~nm) passes through the cell along
$\mathbf{\nabla}\!v$, the shear gradient direction and probes the
sample in only one of the gaps. The scattered pattern
corresponding to light scattered in the velocity-vorticity
($\mathbf{v},\mathbf{z}$) plane is digitalized, by means of a CCD
video camera for the frame acquisition. Experiments at imposed
stress (rheology and SALS) are performed with a commercial
rheometer (RS5, Rheometrics), using a transparent Mooney-Couette
cell with gap $e=0.7$~mm and inner radius $R_i=19$~mm. To
investigate the effect of flow on the
($L_\alpha/L_{\scriptscriptstyle 3}$) phase region, the solution
is stirred in order to obtain a macroscopic homogeneous mixture,
before introducing it into the Couette cell and shearing it at a
constant stress.

\begin{figure}[htbp]
\begin{center}
\scalebox{0.7}{\includegraphics{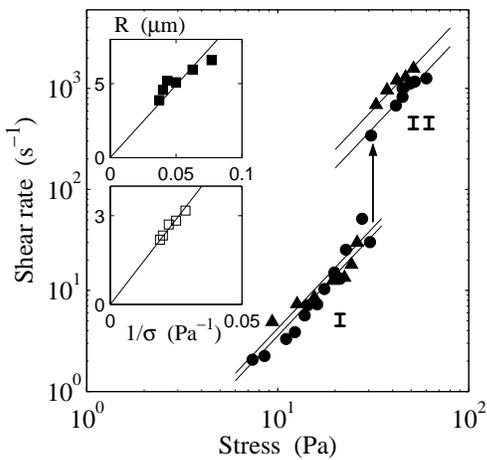}}
\end{center}
\caption{\label{rheologie} Steady shear rate $\gammap$
\textit{vs.} stress $\sigma$ for $T=25^\circ$C and
$S=17$~g.l$^{-1}$ ($\bullet$) and $S=17.15$~g.l$^{-1}$
($\blacktriangle$), the equation of the solid lines are
(respectively for $S=17$~g.l$^{-1}$ and $S=17.15$~g.l$^{-1}$):
$\gammap = 0.042\sigma^2$ and $\gammap = 0.036\sigma^2$ in region
I, $\gammap = 0.410\sigma^2$ and $\gammap = 0.620\sigma^2$ in
region II. For both steady states ($S=17$~g.l$^{-1}$ and
$S=17.15$~g.l$^{-1}$), $\eta = A_{i}\gammap^{-1/2}$ with $\AsA =
6.05$~Pa.s$^{1/2}$ and $\AsA = 5.18$~Pa.s$^{1/2}$, $\AsB =
1.73$~Pa.s$^{1/2}$ and $\AsB = 1.33$~Pa.s$^{1/2}$. Insets: shown
are the steady droplet sizes ($S=17.15$~g.l$^{-1}$) for region I
($\blacksquare$) and II ($\square$) versus $1/\sigma$. Solid lines
are linear fits $R \propto 1/\sigma$}
\end{figure}

\begin{figure}[htbp]
\begin{center}
\scalebox{0.7}{\includegraphics{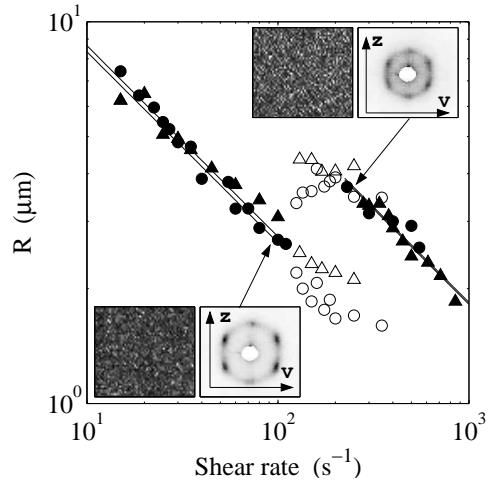}}
\end{center}
\caption{\label{taille} Variation of $R$ the characteristic size
measured from the position of the scattering ring in SALS with
$\gammap$. Circles and triangles correspond respectively to
$S=17$~g.l$^{-1}$ and $S=17.15$~g.l$^{-1}$. Closed
($\blacktriangle$,$\bullet$) symbols represent steady states and
open ($\triangle$,$\circ$) symbols stand for the maxima and minima
of the size during oscillations. The solid straight lines
represent the best fits for $S=17$~g.l$^{-1}$ and
$S=17.15$~g.l$^{-1}$, having slopes: $-1/2$. Inset: shown are the
textures observed between crossed polarizers at $\gammap = 100$
and $\gammap = 230$~s$^{-1}$ and the corresponding SALS patterns.}
\end{figure}

\textit{Results}: After a transient regime which is a few hours
long, a steady state (\textit{i.e.} $\gammap$ does not any longer
vary) is reached (see Fig.~\ref{rheologie}). This steady state I
is characterized in SALS by a scattering ring with a well defined
six fold modulation of its radial intensity (see
Fig.~\ref{taille}). It does not depend on initial conditions and
is determined only by the value of $\sigma$, the applied stress.
If the flow is stopped, the ring persists during a few tens of
minutes with same size, indicating that this shear-induced
structure is metastable. Microscopic observations reveal that this
homogeneous structure consists of a monodisperse closed-compact
multilamellar droplets \cite{Courbin:03} (Inset
Fig.~\ref{taille}).
%%%%%%%%%%%%%%
%%%%%%%%%%%%%%
The ring corresponds to the characteristic size of these closed
packed droplets and its modulation shows the existence of a
hexagonal order in the positions of the droplets. When the stress
is increased, the value of the steady shear rate increases
accordingly to $\gammap \propto \sigma^2$ (indicating a
shear-thinning of the sample since the viscosity $\eta$ varies as
$\eta = A_{\scriptscriptstyle I} \gammap^{-1/2}$, see
Fig.~\ref{rheologie}). The size of the ring becomes larger and
larger indicating that the steady droplet size $R$ decreases. At a
well defined stress $\sigma_c \approx 30$~Pa, a transition
characterized by a jump in the measured shear rate occurs between
steady state I and another steady state, labelled II. Above this
transition, the scattering pattern consists also of a ring with
six well defined spots and the viscosity varies as $\eta =
A_{\scriptscriptstyle II} \gammap^{-1/2}$. Although both steady
states have similar rheological and microscopic features (see
inset of Fig.~\ref{taille}), they can however be easily
distinguished since the scattering ring and therefore the
characteristic droplet size vary discontinuously at the transition
(Fig.~\ref{taille}).
%%%%%%%%%%%%%%
%%%%%%%%%%%%%%

\begin{figure}[htbp]
\begin{center}
\scalebox{0.6}{\includegraphics{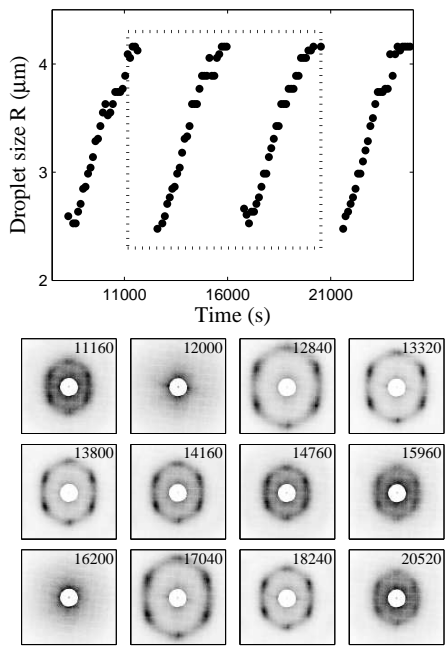}}
\end{center}
\caption{\label{oscillations}
Temporal evolution of $R(t)$, for $S=17.15$~g.l$^{-1}$ and $\gammap = 130$~s$^{-1}$.
Shown is the temporal evolution of the scattering pattern in the ($\mathbf{v},\mathbf{z}$) plane.
}
\end{figure}

To study the nature of the transition between states I and II, we
now perform the same experiment at fixed shear rate. Figure 2
shows the evolution of the steady droplet size $R$, deduced from
SALS measurements, as a function of $\gammap$. Two stationary
branches, labelled I and II are observed for respectively $\gammap
\leq 100$~s$^{-1}$ and  $\gammap \geq 300$~s$^{-1}$. They
correspond to the two steady states already observed in rheology
(see Fig.~\ref{rheologie}). For intermediate shear rates
(\textit{i.e.} $100 \leq \gammap \leq 300$~s$^{-1}$), very
surprisingly, we do not observe any longer steady states, but
instead \textit{oscillatory states}. The size of the scattering
ring oscillates indefinitely in time, indicating that the size of
the droplets becomes a \textit{periodic function} of time
(Fig.~\ref{oscillations}).
%%%%%%%%%%%%%%
%%%%%%%%%%%%%%

\begin{figure}[htbp]
\begin{center}
\scalebox{0.6}{\includegraphics{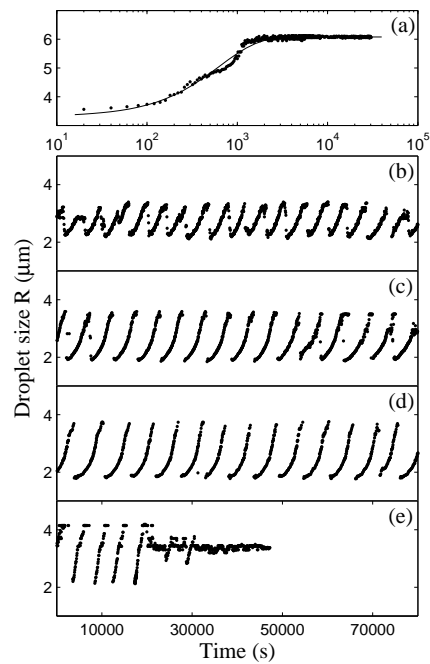}}
\end{center}
\caption{\label{scenario} Temporal evolution of $R(t)$, for (a)
$\gammap = 23$~s$^{-1}$ from an initial shear rate of $\gammap =
54$~s$^{-1}$, the solid line correspond to the best fit using an
exponential expression: $R = 6.08 -2.75 \exp(-t/650)$, (b)
$\gammap = 125$~s$^{-1}$, (c) $\gammap = 150$~s$^{-1}$, (d)
$\gammap = 175$~s$^{-1}$ and (e) to $\gammap = 250$~s$^{-1}$ for
$t<20000$~s and $\gammap = 280$~s$^{-1}$ for $t>20000$~s.}

\end{figure}

Although these data have been recorded and digitalized for one day
($\approx 8\,10^4$~s), we have observed them visually over more
than a week, with no amplitude and period shifts during this time.
The droplet size increases very slowly (the typical duration of
the growth is a few thousands of seconds long) and bursts out
suddenly (in a few tens of seconds). During the continuous growth
of the droplets, no spatial inhomogeneities, such as bands, are
observed in the ($\mathbf{v},\mathbf{z}$) planes. When the laser
beam probing the structure, is moved along the vertical
$\mathbf{z}$--direction, the size of the scattering ring (and
therefore the droplets size) remains constant in the whole cell.
During the burst out of the droplets, the solution becomes turbid,
the scattering ring vanishes and horizontal bands are then
observed in the Couette cell. These reproducible oscillations are
observed both upon increasing or decreasing the shear rate, for
different parameters of the Couette cell (gaps and radii). The
transition I-Oscillation occurs at a well defined shear rate
$\gammap_c \approx 100$~s$^{-1}$ whereas the transition between
Oscillation and II presents an hysteretic behavior for $230 \leq
\gammap \leq 300$~s$^{-1}$ (Fig.~\ref{taille}). Their period (a
few thousands seconds) is many order of magnitude larger than the
period of rotation of the rotor (a few seconds) and varies with
$\gammap$ (see Fig.~\ref{scenario}). These observations rule out
any artefacts, such as a possible coupling with an external
perturbation (building vibrations, temperature fluctuations...).
They clearly prove that the oscillations correspond to a
non-linear relaxation process.
%%%%%%%%%%%%%%
%%%%%%%%%%%%%%

\textit{Discussion}: These observations are very different from
most unsteady flows reported in complex fluids, since they involve
large relaxation oscillations between two different structural
states (I and II) \cite{Remark1}:  the droplet size oscillates. We
believe that a better understanding of the transition I/II may
provide us some insights to the oscillation mechanism. For both
steady states, $R$ varies as $\gammap^{-1/2}$ whereas $\eta
\propto \gammap^{-1/2}$. These two scaling laws suggest that the
droplet size results in both states, from a balance between
surface tension and viscous stress, yielding to the well-known
relation \cite{Taylor:34}:
\begin{equation}
R \approx \frac{\chi}{\eta\gammap} = \frac{\chi}{\sigma} \,.
\label{scaling_taille}
\end{equation}
Within this picture, the surface tensions $\chi$ can be obtained
from the slopes of the linear fits $R$ \textit{vs.} $1/\sigma$
display in the insets of Fig.~\ref{rheologie}. This estimation
respectively leads to $\chi_{\scriptscriptstyle I} \approx 9.5
10^{-5}$~N.m$^{-1}$ for state I and $\chi_{\scriptscriptstyle II}
\approx 11.5 10^{-5}$~N.m$^{-1}$ for state II. Both values are
compatible with what is expected from a dimensional analysis,
namely $\chi \approx \kb T / l^2$, where $\kb T$ is the thermal
energy and $l$ a correlation length of the mixture of the order of
a few hundreds of angstroms, (typically the smectic distance in
the phase $L_\alpha$ or the pore size of the phase
$L_{\scriptscriptstyle 3}$). The transition I/II seems therefore
related to a discontinuity in the surface tension between both
coexisting phases. We believe that a compression of the
multilamellar droplets occurs at this transition yielding to
release of the solvent outside, and as a consequence to the change
 in $\chi$. Such a phenomenon which has already been
evidenced in a $L_\alpha$ phase \cite{Sierro:97} is compatible
with the discontinuous jumps of the viscosity and the size
observed at the transition I/II. It may also explain the turbidity
enhancement occurring when the droplets burst. Within this
picture, there is a critical pressure $\sigma_c$, above which the
droplets compress and release water. More insights on such
mechanism can be gained by considering for instance the following
differential equation, describing a first-order phase transition
induced by Laplace pressure between region I (uncompressed state,
$\phi \approx 0$) and region II (compressed droplets, $\phi\approx
1$):
\begin{align}
\dot{\phi} = -\frac{1}{\taua} \left( \frac{\chi(\phi)/R -
\sigma_c}{\sigma_c} - \alpha(2\phi-1) +(2\phi-1)^3 \right) \,.
\end{align}
Next to this transition, if the shear rate is increased, the size
$R$ slightly decreases (since $R\propto\gammap^{-1/2}$). Then, the
Laplace pressure $\chi/R$ becomes higher than $\sigma_c$ and water
is expelled from the droplets. This water release decreases the
viscosity and consequently the viscous stress. Due to
Eq.~(\ref{scaling_taille}), the droplet size then increases again
and the Laplace pressure decreases. As a result, the water goes
back into the droplets and the same mechanism starts over leading
to sustained oscillations. Because permeation of water through
membranes is a very slow process \cite{Leng:01}, such a scenario
may explain the very large time scales involved in our
observations, as suggested in Refs.~\cite{Wunenburger:01}.

In a first approach, for simplicity's sake, we consider that the
temporal evolution of $R(t)$ is driven by the following
first-order kinetics:
\begin{align}
 \dot{R} = -\frac{1}{\taub} \left( R -
\frac{\chi(\phi)}{\eta(\phi,\gammap)\gammap} \right) \,,
\label{model}
\end{align}
with $\chi(\phi) = \chiA(1-\phi) + \chiB \phi$, and
$\eta(\phi,\gammap) = A(\phi)\gammap^{-1/2}$ where  $A(\phi) =
\AsA(1-\phi) + \AsB \phi$. Taking for $\AsA$, $\AsB$, $\chiA$,
$\chiB$, $\sigma_c$ the values found experimentally
(Figs.~\ref{rheologie} and \ref{taille}) and $\taub = 650$~s
(Fig.~\ref{scenario}), our simplified two equations reproduce the
two steady branches observed experimentally for both the droplet
size and the viscosity. For intermediate shear rates, in a given
range of the $\alpha$ and $\taua$ parameters, it may also lead to
sustained droplet size oscillations \cite{Remark2}. The transient
banding behavior observed during the fast burst of the droplets
suggests that an overall description requires the set-up of space
and time-dependent non-linear coupled differential equations. Such
approaches are currently being developed with success by theorists
to describe spatio-temporal flows of complex fluids
\cite{Head:01,Picard:02}. We hope that our experiments on a model
system will stimulate such theoretical researches.

To conclude, we have shown in a two phase flow, the existence of
\textit{non-linear structural oscillatory states} for which
droplet size is a \textit{periodic function} of time. This new
result raises now an important question: can the droplet size
display \textit{chaotic} behavior? We believe that this question
will be a major topic of research in the study of
\textit{rheochaos} \cite{Head:01} in the near future.

\acknowledgments{ We thank T. Douar and M. Winckert for technical
assistance. We acknowledge fruitful discussions with A. Benayad,
A. Colin, J.-P. Delville and J. Rouch.}


\begin{thebibliography}{22}
\expandafter\ifx\csname natexlab\endcsname\relax\def\natexlab#1{#1}\fi
\expandafter\ifx\csname bibnamefont\endcsname\relax
  \def\bibnamefont#1{#1}\fi
\expandafter\ifx\csname bibfnamefont\endcsname\relax
  \def\bibfnamefont#1{#1}\fi
\expandafter\ifx\csname citenamefont\endcsname\relax
  \def\citenamefont#1{#1}\fi
\expandafter\ifx\csname url\endcsname\relax
  \def\url#1{\texttt{#1}}\fi
\expandafter\ifx\csname urlprefix\endcsname\relax\def\urlprefix{URL }\fi
\providecommand{\bibinfo}[2]{#2}
\providecommand{\eprint}[2][]{\url{#2}}

\bibitem[{\citenamefont{Larson}(1999)}]{Larson:99}
\bibinfo{author}{\bibfnamefont{R.~G.} \bibnamefont{Larson}},
  \emph{\bibinfo{title}{The Structure and Rheology of Complex Fluids}}
  (\bibinfo{publisher}{Oxford University Press}, \bibinfo{year}{1999}).

\bibitem[{\citenamefont{Onuki}(1997)}]{Onuki:97}
\bibinfo{author}{\bibfnamefont{A.}~\bibnamefont{Onuki}}, \bibinfo{journal}{J.
  Phys.: Condens. Matter} \textbf{\bibinfo{volume}{9}}, \bibinfo{pages}{6119}
  (\bibinfo{year}{1997}).

\bibitem[{\citenamefont{Taylor}(1934)}]{Taylor:34}
\bibinfo{author}{\bibfnamefont{G.~I.} \bibnamefont{Taylor}},
  \bibinfo{journal}{Proc. R. Soc. London, Ser. A}
  \textbf{\bibinfo{volume}{146}}, \bibinfo{pages}{501} (\bibinfo{year}{1934}).

\bibitem[{\citenamefont{Hashimoto et~al.}(1995)\citenamefont{Hashimoto,
  Matsuzaka, and Moses}}]{Hashimoto:95}
\bibinfo{author}{\bibfnamefont{T.}~\bibnamefont{Hashimoto}},
  \bibinfo{author}{\bibfnamefont{K.}~\bibnamefont{Matsuzaka}},
  \bibnamefont{and} \bibinfo{author}{\bibfnamefont{E.}~\bibnamefont{Moses}},
  \bibinfo{journal}{Phys. Rev. Lett.} \textbf{\bibinfo{volume}{74}},
  \bibinfo{pages}{126} (\bibinfo{year}{1995}).

\bibitem[{\citenamefont{Min et~al.}(1989)\citenamefont{Min, Stavans, Piazza,
  and Goldburg}}]{Min:89}
\bibinfo{author}{\bibfnamefont{K.~Y.} \bibnamefont{Min}},
  \bibinfo{author}{\bibfnamefont{J.}~\bibnamefont{Stavans}},
  \bibinfo{author}{\bibfnamefont{R.}~\bibnamefont{Piazza}}, \bibnamefont{and}
  \bibinfo{author}{\bibfnamefont{W.~I.} \bibnamefont{Goldburg}},
  \bibinfo{journal}{Phys. Rev. Lett.} \textbf{\bibinfo{volume}{63}},
  \bibinfo{pages}{1070} (\bibinfo{year}{1989}).

\bibitem[{\citenamefont{Cristobal et~al.}(2000)\citenamefont{Cristobal, Rouch,
  Colin, Panizza, and Narayanan}}]{Cristobal:00}
\bibinfo{author}{\bibfnamefont{G.}~\bibnamefont{Cristobal}},
  \bibinfo{author}{\bibfnamefont{J.}~\bibnamefont{Rouch}},
  \bibinfo{author}{\bibfnamefont{A.}~\bibnamefont{Colin}},
  \bibinfo{author}{\bibfnamefont{P.}~\bibnamefont{Panizza}}, \bibnamefont{and}
  \bibinfo{author}{\bibfnamefont{T.}~\bibnamefont{Narayanan}},
  \bibinfo{journal}{Phys. Rev. E} \textbf{\bibinfo{volume}{62}},
  \bibinfo{pages}{3871} (\bibinfo{year}{2000}); \bibinfo{journal}{Phys. Rev. E} \textbf{\bibinfo{volume}{64}},
  \bibinfo{pages}{011505}(\bibinfo{year}{2001}).

\bibitem[{\citenamefont{Mason and Bibette}(1996)}]{Mason:96}
\bibinfo{author}{\bibfnamefont{T.~G.} \bibnamefont{Mason}} \bibnamefont{and}
  \bibinfo{author}{\bibfnamefont{J.}~\bibnamefont{Bibette}},
  \bibinfo{journal}{Phys. Rev. Lett.} \textbf{\bibinfo{volume}{77}},
  \bibinfo{pages}{3481} (\bibinfo{year}{1996}).

\bibitem[{\citenamefont{Bonn et~al.}(1993)\citenamefont{Bonn, Meunier,
Greffier, Al-Kahwaji, and Kellay}}]{Bonn:98}
  \bibinfo{author}{\bibfnamefont{D.} \bibnamefont{Bonn}},
  \bibinfo{author}{\bibfnamefont{J.}~\bibnamefont{Meunier}},
  \bibinfo{author}{\bibfnamefont{O.}~\bibnamefont{Greffier}},
  \bibinfo{author}{\bibfnamefont{A.}~\bibnamefont{Al-Kahwaji}}
  \bibnamefont{and} \bibinfo{author}{\bibfnamefont{H.}~\bibnamefont{Kellay}},
 \bibinfo{journal}{Phys. Rev. E} \textbf{\bibinfo{volume}{58}},
  \bibinfo{pages}{2115} (\bibinfo{year}{1998}) and refs. therein.

\bibitem[{\citenamefont{Olmsted and Lu}(1997)}]{Olmsted:97}
\bibinfo{author}{\bibfnamefont{P.~D.} \bibnamefont{Olmsted}} \bibnamefont{and}
  \bibinfo{author}{\bibfnamefont{C.-Y.~D.} \bibnamefont{Lu}},
  \bibinfo{journal}{Phys. Rev. E} \textbf{\bibinfo{volume}{56}},
  \bibinfo{pages}{55} (\bibinfo{year}{1997}); \bibinfo{author}{\bibfnamefont{P.~D.} \bibnamefont{Olmsted}}
\bibinfo{journal}{Europhys. Lett.} \textbf{\bibinfo{volume}{48}},
  \bibinfo{pages}{339} (\bibinfo{year}{1999}) and refs. therein.

\bibitem[{\citenamefont{Wunenburger et~al.}(2001)\citenamefont{Wunenburger,
  Colin, Leng, Arn\'{e}odo, and Roux}}]{Wunenburger:01}
\bibinfo{author}{\bibfnamefont{A.-S.} \bibnamefont{Wunenburger}},
  \bibinfo{author}{\bibfnamefont{A.}~\bibnamefont{Colin}},
  \bibinfo{author}{\bibfnamefont{J.}~\bibnamefont{Leng}},
  \bibinfo{author}{\bibfnamefont{A.}~\bibnamefont{Arn\'{e}odo}},
  \bibnamefont{and} \bibinfo{author}{\bibfnamefont{D.}~\bibnamefont{Roux}},
  \bibinfo{journal}{Phys. Rev. Lett.} \textbf{\bibinfo{volume}{86}},
  \bibinfo{pages}{1374} (\bibinfo{year}{2001});
  \bibinfo{author}{\bibfnamefont{J.-B.} \bibnamefont{Salmon}},
  \bibinfo{author}{\bibfnamefont{A.}~\bibnamefont{Colin}}
  \bibnamefont{and} \bibinfo{author}{\bibfnamefont{D.}~\bibnamefont{Roux}},
 \bibinfo{journal}{Phys. Rev. E} \textbf{\bibinfo{volume}{66}},
  \bibinfo{pages}{031505} (\bibinfo{year}{2002}).

\bibitem[{\citenamefont{Bandyopadhyay et~al.}(2000)\citenamefont{Bandyopadhyay,
  Basappa, and Sood}}]{Bandyopadhyay:00}
\bibinfo{author}{\bibfnamefont{R.}~\bibnamefont{Bandyopadhyay}},
  \bibinfo{author}{\bibfnamefont{G.}~\bibnamefont{Basappa}}, \bibnamefont{and}
  \bibinfo{author}{\bibfnamefont{A.~K.} \bibnamefont{Sood}},
  \bibinfo{journal}{Phys. Rev. Lett.} \textbf{\bibinfo{volume}{84}},
  \bibinfo{pages}{2022} (\bibinfo{year}{2000}).

\bibitem[{\citenamefont{Head et~al.}(2001)\citenamefont{Head, Ajdari, and
  Cates}}]{Head:01}
\bibinfo{author}{\bibfnamefont{D.~A.} \bibnamefont{Head}},
  \bibinfo{author}{\bibfnamefont{A.}~\bibnamefont{Ajdari}}, \bibnamefont{and}
  \bibinfo{author}{\bibfnamefont{M.~E.} \bibnamefont{Cates}},
  \bibinfo{journal}{Phys. Rev. E} \textbf{\bibinfo{volume}{64}},
  \bibinfo{pages}{061509} (\bibinfo{year}{2001});
  \bibinfo{journal}{Europhys. Lett.} \textbf{\bibinfo{volume}{57}},
  \bibinfo{pages}{120} (\bibinfo{year}{2001}).

\bibitem[{\citenamefont{Picard et~al.}(2002)\citenamefont{Picard, Ajdari,
  Bocquet, and Lequeux}}]{Picard:02}
\bibinfo{author}{\bibfnamefont{G.}~\bibnamefont{Picard}},
  \bibinfo{author}{\bibfnamefont{A.}~\bibnamefont{Ajdari}},
  \bibinfo{author}{\bibfnamefont{L.}~\bibnamefont{Bocquet}}, \bibnamefont{and}
  \bibinfo{author}{\bibfnamefont{F.}~\bibnamefont{Lequeux}},
  \bibinfo{journal}{Phys. Rev. E} \textbf{\bibinfo{volume}{66}},
  \bibinfo{pages}{051501} (\bibinfo{year}{2002}).

\bibitem[{\citenamefont{Grosso et~al.}(2001)\citenamefont{Grosso, Keunings,
  Crescitelli, and Maffettone}}]{Grosso:01}
\bibinfo{author}{\bibfnamefont{M.}~\bibnamefont{Grosso}},
  \bibinfo{author}{\bibfnamefont{R.}~\bibnamefont{Keunings}},
  \bibinfo{author}{\bibfnamefont{S.}~\bibnamefont{Crescitelli}},
  \bibnamefont{and} \bibinfo{author}{\bibfnamefont{P.~L.}
  \bibnamefont{Maffettone}}, \bibinfo{journal}{Phys. Rev. Lett.}
  \textbf{\bibinfo{volume}{86}}, \bibinfo{pages}{3184} (\bibinfo{year}{2001}).

\bibitem[{\citenamefont{Gosh and Miller}(1987)}]{Gosh:87}
\bibinfo{author}{\bibfnamefont{O.}~\bibnamefont{Gosh}} \bibnamefont{and}
  \bibinfo{author}{\bibfnamefont{C.~A.} \bibnamefont{Miller}},
  \bibinfo{journal}{J. Phys. Chem.} \textbf{\bibinfo{volume}{91}},
  \bibinfo{pages}{258} (\bibinfo{year}{1987}).

\bibitem[{\citenamefont{Porte}(1992)}]{Porte:92}
\bibinfo{author}{\bibfnamefont{G.}~\bibnamefont{Porte}}, \bibinfo{journal}{J.
  Phys.: Condens. Matter} \textbf{\bibinfo{volume}{4}}, \bibinfo{pages}{8649}
  (\bibinfo{year}{1992}).

\bibitem[{\citenamefont{Courbin et~al.}(2002)\citenamefont{Courbin, Delville,
  Rouch, and Panizza}}]{Courbin:02}
\bibinfo{author}{\bibfnamefont{L.}~\bibnamefont{Courbin}},
  \bibinfo{author}{\bibfnamefont{J.-P.} \bibnamefont{Delville}},
  \bibinfo{author}{\bibfnamefont{J.}~\bibnamefont{Rouch}}, \bibnamefont{and}
  \bibinfo{author}{\bibfnamefont{P.}~\bibnamefont{Panizza}},
  \bibinfo{journal}{Phys. Rev. Lett.} \textbf{\bibinfo{volume}{89}},
  \bibinfo{pages}{148305} (\bibinfo{year}{2002}).

\bibitem[{\citenamefont{Courbin et~al.}(2003)\citenamefont{Courbin, Pons, Rouch, and Panizza}}]{Courbin:03}
\bibinfo{author}{\bibfnamefont{L.}~\bibnamefont{Courbin}},
  \bibinfo{author}{\bibfnamefont{R.}~\bibnamefont{Pons}},
  \bibinfo{author}{\bibfnamefont{J.}~\bibnamefont{Rouch}}, \bibnamefont{and}
  \bibinfo{author}{\bibfnamefont{P.}~\bibnamefont{Panizza}},
  \bibinfo{journal}{Europhys. Lett.} \textbf{\bibinfo{volume}{61}},
  \bibinfo{pages}{275} (\bibinfo{year}{2003}).

\bibitem[{Rem()}]{Remark1}
\bibinfo{note}{Structural oscillations between a disordered and an ordered
  state have been reported by Wunenburger \textit{et al.} in a $L_\alpha$ phase
  at fixed stress \cite{Wunenburger:01}}.

\bibitem[{\citenamefont{Sierro and Roux}(1997)}]{Sierro:97}
\bibinfo{author}{\bibfnamefont{O.} \bibnamefont{Diat}},
\bibinfo{author}{\bibfnamefont{D.} \bibnamefont{Roux}} \bibnamefont{and}
  \bibinfo{author}{\bibfnamefont{F.} \bibnamefont{Nallet}},
  \bibinfo{journal}{Phys. Rev. E} \textbf{\bibinfo{volume}{51}},
  \bibinfo{pages}{3296} (\bibinfo{year}{1995});
\bibinfo{author}{\bibfnamefont{P.}~\bibnamefont{Sierro}} \bibnamefont{and}
  \bibinfo{author}{\bibfnamefont{D.}~\bibnamefont{Roux}},
  \bibinfo{journal}{Phys. Rev. Lett.} \textbf{\bibinfo{volume}{78}},
  \bibinfo{pages}{1496} (\bibinfo{year}{1997}).

\bibitem[{\citenamefont{Leng and Nallet and Roux }(2001)}]{Leng:01}
\bibinfo{author}{\bibfnamefont{J.}~\bibnamefont{Leng}},
  \bibinfo{author}{\bibfnamefont{F.}~\bibnamefont{Nallet}}, \bibnamefont{and}
\bibinfo{author}{\bibfnamefont{D.}~\bibnamefont{Roux}},
  \bibinfo{journal}{Eur. Phys. J. E} \textbf{\bibinfo{volume}{4}},
  \bibinfo{pages}{337} (\bibinfo{year}{2001}).



\bibitem[{\citenamefont{Courbin et~al.}()\citenamefont{Courbin, Panizza, and
  Salmon}}]{Remark2}
\bibinfo{author}{\bibfnamefont{L.}~\bibnamefont{Courbin}},
  \bibinfo{author}{\bibfnamefont{P.}~\bibnamefont{Panizza}}, \bibnamefont{and}
  \bibinfo{author}{\bibfnamefont{J.-B.} \bibnamefont{Salmon}},
  \bibinfo{note}{to be published}.

\end{thebibliography}
\end{document}